\documentclass[12pt]{article}
\usepackage{amsmath,amssymb,latexsym}
\usepackage[dvips]{graphicx}
\providecommand{\beqa}{\begin{eqnarray}}

\providecommand{\eeqa}{\end{eqnarray}}
\providecommand{\ti}{{\mathrm{t}}}

\def\Orb{{\mathbf{S}^1/\mathbf{Z}_2}}
\def\Z2{{\mathbf{Z}_2}}
\def\mX{{\mathbf{X}}}

\begin{document}

\thispagestyle{empty}

\addtolength{\baselineskip}{1.5mm}

\thispagestyle{empty}

\vspace{28pt}

\begin{center}
{\Large\textbf{Energy Transfer in Multi Field Inflation and\\ \vskip2mm Cosmological Perturbations}}\\

\vspace{28pt}

Amjad Ashoorioon$^\blacklozenge$\footnote{amjad@umich.edu}, Axel
Krause$^\clubsuit$\footnote{axel.krause@physik.uni-muenchen.de} and
Krzysztof
Turzynski$^{\blacklozenge\spadesuit}$\footnote{turzyn@fuw.edu.pl}

\vspace{24pt}

${}^\blacklozenge$\textit{Michigan Center for Theoretical Physics\\
University of Michigan, Ann Arbor\\
Michigan 48109-1040, USA}\\ \vskip2mm

${}^\clubsuit$\textit{Arnold Sommerfeld Center for Theoretical
Physics\\
Department f\"{u}r Physik, Ludwig-Maximilians-Universit\"{a}t M\"{u}nchen,\\
Theresienstr.~37, 80333 Munich, Germany}\\ \vskip2mm

${}^\spadesuit$\textit{Institute of Theoretical Physics\\
University of Warsaw,\\
ul.~Ho\. za 69, 00-681 Warsaw, Poland}\\

\end{center}

\vspace{18pt}

\begin{abstract}

\addtolength{\baselineskip}{1.2mm}

In cascade inflation and some other string inflation models,
collisions of mobile branes with other branes or orbifold planes
occur and lead to interesting cosmological signatures. The
fundamental M/string-theory description of these collisions is still
lacking but it is clear that the inflaton looses part of its energy
to some form of brane matter, e.g.~a component of tensionless
strings. In the absence of a fundamental description, we assume a
general barotropic fluid on the brane, which absorbs part of the
inflaton's energy. The fluid is modeled by a scalar with a suitable
exponential potential to arrive at a full-fledged field theory
model. We study numerically the impact of the energy transfer from the
inflaton to the scalar on curvature and isocurvature
perturbations and demonstrate explicitly that the curvature power
spectrum gets modulated by oscillations which damp away toward
smaller scales. Even though, the contribution of isocurvature
perturbations decays toward the end of inflation, they induce
curvature perturbations on scales that exit the horizon before the
collision. We consider cases where the scalar behaves like
radiation, matter or a web of cosmic strings and discuss the
differences in the resulting power spectra.

\end{abstract}

\setcounter{footnote}{0}

\newpage

\section{Introduction}

There has been a lot of activity over the past few years to derive
inflation from string theory. Most prominent among the many
approaches has been brane inflation, in which the coordinates of
some mobile branes on the internal compactification manifold provide
the inflatons in the effective four-dimensional low energy theory
\cite{Dvali:1998pa}. While initial efforts focussed, for simplicity,
on compactifications with a single mobile brane (for recent updates
on single brane inflations, see
e.g.~\cite{Kachru:2003sx}-\cite{Hoi:2008gc}) general
compactifications with fluxes naturally possess several, often many,
mobile branes to satisfy tadpole cancellation equations
\cite{Becker:2005sg}-\cite{Ward:2007gs}. Multi brane inflation
models lead to multi field inflation cosmologies in the low energy
theory, which as a result have been actively researched recently
(see e.g.~\cite{Dimopoulos:2005ac}-\cite{Cai:2008if}). The multitude
of inflaton fields can be actually seen as a physical blessing
rather than a technical curse.

The multi brane inflation models split into two classes: those with
brane-brane interactions which steer the multi brane system towards
a stable dynamical attractor and those with brane-brane interactions
for which no such attractor exists. The first class allows to
replace the multi field cosmologies by cosmologies involving fewer
fields, in the extreme case just a single inflaton field, once the
system evolves along the attractor. An example for the first class
is a multi brane system with exponential interactions among the
branes, caused by non-perturbative interactions between classically
non-interacting branes \cite{Becker:2005sg}. As a result of the
stable attractor evolution this class more readily allows for an
analytical treatment and moreover offers a parametric way to achieve
inflation. Namely by choosing the number of participating branes
large enough, one achieves a parametric suppression of the slow-roll
parameters -- a phenomenon which is known as assisted inflation
\cite{Liddle:1998jc}.

The second class, on the other hand, requires an enormous
fine-tuning for the position of each mobile brane in order to yield
an ordered, e.g.~equidistant, brane configuration, which admits an
analytical treatment. An example of this class would be a system
consisting of many brane-antibrane pairs with mutual Coulomb
interactions. Full-fledged numerical studies for the cosmic
evolution of this second class still need to be performed.
Nevertheless, one might expect that this second class won't lead to
inflation. The reason is that even if one starts with a multi brane
configuration suitable for inflation (which, as said, requires a
tremendous fine-tuning for all brane positions as the system lacks
an attractor), small perturbations won't die away. They will lead
the system to a complicated multi brane dynamics which moves the
branes around on the internal space in an unconcerted way,
precluding the realization of assisted inflation. The upshot might
be that among the many possible multi brane models, the models of
the first class are the ones capable of generating inflation,
whereas the models of the second class have to be dismissed.

In this paper we focus on a new phenomenon which arises in multi
brane inflation models: the possibility of cascade inflation phases
\cite{Ashoorioon:2006wc}. These arise when some of the mobile branes
collide successively with fixed branes or orbifold fixed planes and
thus no longer participate in the inflation process. The parameters
of the effective four-dimensional inflationary potential vary in
these collisions and as a result the inflationary potential acquires
features. A characteristic consequence of such features is a damped
oscillatory behavior of the power spectrum of density perturbations.
This could lead to important observational clues about the
underlying time variation of the inflationary potential.

To date, we have no complete fundamental M/string-theory description
of such collisions. Therefore, we have to model the collision in
field theory as an energy transfer (ET) from the inflaton to the
brane. For instance, the brane could be modeled by a perfect fluid
with a coupling to the inflaton which allows to transfer the
inflaton's energy in discrete steps. More adapt to a field theory
description based on a Lagrangean, and the route we follow here, is
to model the perfect fluid by a scalar field with appropriate
potential. Our goal in this paper is to study the potentially
observable imprints of this ET on the resulting curvature and
isocurvature perturbation spectrum. Chapter \ref{chapter-cascade}
describes the M-theory motivation and background for our analysis,
reviewing cascade inflation. Chapter \ref{chapter-Model} introduces
our two scalar field theory, which models the ET in brane collisions
such as arising in cascade inflation. Chapter \ref{chapter-Spectra}
presents the curvature and isocurvature perturbations for this
model. We find that the contribution of isocurvature perturbations
decays toward the end of inflation. Nonetheless, they induce
curvature perturbations on scales that exit the horizon before the
collision. The curvature power spectrum gets modulated by
oscillations which damp away toward smaller scales. Finally,
technical details on how to calculate the perturbations in our
two-field model are presented in the Appendix.

\section{Cascade Inflation}\label{chapter-cascade}

The goal of this paper is to study perturbations generated in a two
scalar field inflation model with exponential potentials. Such
potentials arise in fundamental physics from M-theory multi brane
inflation  \cite{Becker:2005sg} and cascade inflation
\cite{Ashoorioon:2006wc} through non-perturbative instanton
interactions. Since the two scalar field inflation model might also
be used to provide some first insight into the cosmological impact
of the ET arising when a bulk M5-brane collides with the boundary in
cascade inflation, we will now briefly recall cascade inflation as
one of the motivations for our study.

Cascade inflation arises from heterotic M-theory
\cite{Horava:1995qa}, \cite{Horava:1996ma} compactified on
$\mX\times\Orb$, where $\mX$ is a Calabi-Yau threefold. Tadpole
cancellation requires generically the presences of $N$ M5-branes in
the background. These fill the four-dimensional spacetime and wrap a
holomorphic two-cycle on $\mX$ of genus zero. As BPS objects, the
M5-branes interact only non-perturbatively via open M2-instantons
which stretch along $\Orb$ and wrap the same two-cycle\footnote{For
simplicity we are assuming a Calabi-Yau threefold with Hodge number
$h^{(1,1)}(\mX) = 1$.}. The effective four-dimensional ${\cal N}=1$
supergravity resulting from this compactification contains as scalar
components of complex chiral superfields the M5-brane position
moduli $Y_i, \; i=1,\hdots,N$, the Calabi-Yau $\mX$ volume modulus
$S$ and the $\Orb$ orbifold-size modulus $T$. Twice their real parts
are denoted by $y_i$, $s$ and $t$. Moreover, it's convenient to
define $y = (\sum_{i=1}^N y_i^2)^{1/2}$. One can show that the
M-theory dynamics leads to an equidistant distribution of the
M5-branes along $\Orb$ \cite{Becker:2005sg}. For such a distribution
\beqa \text{Re}(Y_{i+1}-Y_i) \equiv \bigg(\frac{t}{2L}\bigg) \Delta
x \eeqa is independent of the M5-brane counting label $i$.
Consequently, the $N-1$ inflatons $\text{Re}(Y_{i+1}-Y_i)$ can be
identified. The multi inflaton system reduces effectively to one
with a single inflaton $\Delta x$. The $\Orb$ interval size is
denoted by $L$.

In the large volume limit, specified by the inequality $st \gg y^2$,
where supergravity provides a reliable description of the dynamics
\cite{Becker:2005sg}, the potential for the canonically normalized
inflaton $\varphi \sim N^{3/2} \Delta x$ becomes
\beqa
V_N(\varphi)
= V_N e^{-\sqrt{\frac{2}{p_N}}\frac{\varphi}{M_{\rm P}}} \; ,
\label{Potential}
\eeqa
where
\beqa V_N = (N-1)^2
\bigg(\frac{6M_{\rm P}^4}{st^3 d}\bigg) \; , \qquad p_N =
\frac{4N(N^2-1)}{3st}
\eeqa
and $d$ is the Calabi-Yau intersection
number. For the detailed derivation from M-theory we refer the
reader to \cite{Becker:2005sg,Ashoorioon:2006wc}. The cosmological
FRW evolution in this background is given by power law inflation
\cite{Lucchin:1984yf} with FRW scale factor \beqa a(\ti) = a_0
\ti^{p_N} \; . \eeqa Inflation sets in when $p_N > 1$, which can
always be achieved when sufficiently many M5-branes are present
since $p_N \sim N^3$. Besides this bound on $N$ from below, there is
also a bound on $N$ from above which follows from the requirement to
work in the large volume regime where $st\gg y^2$ holds and the fact
that $y$ grows with $N$. For typical parameter values one thus finds
$20 \le N \le 200$ as a constraint on the number of M5-branes
\cite{Becker:2005sg}. Such numbers can easily be accounted for in
heterotic M-theory flux compactifications
\cite{Witten:1996mz,Curio:2000dw,Curio:2003ur} where tadpole
cancellation equations balance the amount of M5-branes with
quantized flux numbers.

The repulsive M2-instanton interactions between the M5-branes cause
them to spread over the $\Orb$ interval until the two outermost
M5-branes hit the boundaries and dissolve into them via small
instanton transitions \cite{Witten:1995gx}, \cite{Ovrut:2000qi}, \cite{Buchbinder:2002ji}. This process changes the
topological data on the boundaries while the number of M5-branes
participating in the inflationary bulk dynamics drops from $N$ to
$N-2$. The remaining $N-2$ bulk M5-branes will continue to spread
until again the most outermost M5-branes hit the boundaries in a
second small instanton transition and so on. This evolution, in
which the number of M5-branes drops successively in discrete steps,
defines \emph{cascade inflation} \cite{Ashoorioon:2006wc}. In
\cite{Ashoorioon:2006wc}, the analysis neglected the energy
transferred to the boundaries by the instanton transitions and
therefore worked, after each instanton transition, with a suitably
modified power-law evolution
\begin{equation}\label{scl-fac}
a_{m}(\mathrm{t}) = a_{m}\mathrm{t}^{p_{N_m}}\; , \qquad
{\mathrm{t}}_{m-1}\leq \mathrm{t}\leq \mathrm{t}_m \; ,
\end{equation}
having different $N$ dependent parameters $p_N$ and $V_N$ for each
interval. The cascade inflation process terminates  when the number
of M5-branes in the $m$th phase, $N_m = N-2m$, drops below a
critical value $N_K$ in the final $K$th phase. This critical value
$N_K$ is determined by the exit condition $p_{N_K} = 1$, at which
inflation stops and which is dynamically reached from larger powers
$p_{N_m} \ge p_{N_K}$, $m \le K$. Thus we have a finite number
$m=1,\hdots, K$ of cascade inflation bouts.

Throughout the whole cascade inflation process the inflaton, $\Delta
x$, representing the distance between neighboring bulk M5-branes,
grows continuously. Matching the scale factors at the transition
times, $\mathrm{t}_m$, determines the prefactors \beqa a_{m}= a_1
\mathrm{t}_1^{p_{N_1}-p_{N_2}} \mathrm{t}_2^{p_{N_2}-p_{N_3}} \ldots
\mathrm{t}_{m-1}^{p_{N_{m-1}}-p_{N_m}} \; , \eeqa where
$\mathrm{t}_{ij}=\mathrm{t}_i/\mathrm{t}_j$. The scale factor, but
not the Hubble parameter, is continuous at the transition times
$\ti_m$, when the ET to the boundaries is neglected. The onset time
of inflation, $\mathrm{t}_0$, is determined by inverting the
power-law inflation solution for $\varphi(\ti)$ in the initial phase
and noting that $\Delta x(\mathrm{t}_0)/L\ll 1$. The result is
\begin{equation}\label{t0}
\mathrm{t}_{0} \simeq \frac{2 N^2}{3M_{\rm P}} \sqrt{\frac{2td}{s}} \; .
\end{equation}
Inverting the solution for $\varphi(\ti)$ gives the transition times
\beqa\label{tr-time}
(\mathrm{t}_m - \mathrm{t}_0) M_{\rm P} = \bigg(
\frac{p_{N_1}(3p_{N_1}-1)}{N_1-1} e^{\frac{t}{N_1}}
+ \sum_{k=2}^m
\frac{p_{N_k}(3p_{N_k}-1)}{N_k-1}
e^{\frac{t}{N_k-1}-\frac{t}{N_{k-1}-1}} \bigg)
\sqrt{\frac{st^3 d}{6}} \; ,
\eeqa
from which the number of e-foldings, generated during cascade
inflation, derives
\begin{equation}\label{Ne}
N_e = \ln\left(\frac{a(\mathrm{t}_f)}{a(\mathrm{t}_0)}\right) =
\sum_{m=1}^{K} p_{N_m}\ln (\mathrm{t}_{m,m-1}) \; .
\end{equation}

The analysis of \cite{Ashoorioon:2006wc} neglected the backreaction
of the energy, which gets lost to the boundaries. The reason for
this was the still open question how the small instanton transitions
should be described dynamically at a fundamental level plus the
naive expectation that, as long, as the number of M5-branes, having
been absorbed by the boundaries, stays small compared to those
remaining in the bulk and driving inflation, this might be a useful
approximation. For the type I string, possessing an $SO(32)$ gauge
group, the small $SO(32)$ instanton is nothing but a D5-brane in ten
dimensions \cite{Seiberg:1996vs}. When compactified on a K3 manifold
from ten to six dimensions, a six-dimensional instanton arises from
the D5-brane wrapping the K3 which describes an effective string in
six dimensions (as the string theory situation is best understood in
six dimensions, we stick to this case for this brief discussion). At
a certain point in moduli space the tension of this string vanishes
and a singularity occurs. At this point the string's sigma model
coupling is of order one, hence the sigma model is strongly coupled,
whereas the string coupling constant could be arbitrarily small
\cite{Seiberg:1996vs}. The type I - heterotic duality relates this
type I phenomena to a singularity involving the fundamental $SO(32)$
heterotic string in six dimensions. Furthermore, invoking T-duality,
one can relate the phenomenon further to the $E_8\times E_8$
heterotic string in six dimensions \cite{Ganor:1996mu}, for which at
the singularity one of the two $E_8$ gauge couplings diverges and
the string associated with the small instanton in that gauge group
becomes tensionless, while the other $E_8$ gauge coupling remains
finite.

In heterotic M-theory, compactified on K3$\times\Orb$ from eleven
down to six dimensions, the two $E_8$ gauge groups are geometrically
separated and localized on the ends of the $\Orb$ interval. For
these we can first of all have the same instanton in either $E_8$
gauge theory as for the weakly coupled $E_8\times E_8$ heterotic
string with the associated tensionless string when the $E_8$
instanton shrinks and produces a singularity. However, a novel
situation arises from the presence of M5-branes, which are
generically needed to cancel anomalies. In six dimensions, these
M5-branes are located as points on the compactification manifold
K3$\times\Orb$. An M2-brane stretches along $\Orb$ from the M5-brane
to each boundary. Such an M2-brane produces a string in the
non-compact six dimensions, and generates a tensionless string once
the M5-brane hits a boundary \cite{Ganor:1996mu},
\cite{Seiberg:1996vs}. In heterotic M-theory compactifications on
$\mX\times\Orb$ down to four dimensions, the small instantons are
described by a torsion free sheaf, a singular bundle. The singular
torsion free sheaf can then be smoothed out to a non-singular
holomorphic vector bundle by moving in moduli space
\cite{Ovrut:2000qi}.

To date, unfortunately, no clear fundamental M-theory description of
these small instanton transitions is available, which would fully
describe its dynamics, including the produced tensionless strings.
In what follows, switching back to a four dimensional analysis of
the ensuing cosmology, we will therefore adopt a quantum field
theory description which models such a transition by coupling the
inflaton $\varphi$ to another field $\chi$ (which in heterotic
M-theory would come from the boundary). To allow for an energy
transfer from the inflaton $\varphi$ to the boundary field $\chi$,
we will introduce a suitable coupling between the two fields.

\section{The Two Field Model}\label{chapter-Model}

\subsection{The Potential}

As mentioned above, repulsive M2-instanton interactions cause the
M5-branes to spread over the $\Orb$ interval until the two outermost
M5-branes hit the interval's boundaries. The ensuing
non-perturbative small instanton transition transforms the outermost
M5-branes into small instantons on the boundaries
\cite{Witten:1995gx}. This process changes the topological data on
the boundaries while the number $N$ of M5-branes, participating in
the inflationary bulk dynamics, drops to $N-2$ at each such
transition. The resulting features in the inflaton's potential lead
to interesting new observational phenomena. These have been derived
in \cite{Ashoorioon:2006wc} under neglect of the full backreaction
of the energy which gets transferred to the boundaries. To get a
first clue how this backreaction alters the observational signatures of \cite{Ashoorioon:2006wc}, we will now introduce a simple two field model which allows us to study the backreaction effects numerically. In our subsequent analysis we will restrict ourselves to a single collision, i.e.~a single step in the inflaton potential. Furthermore, to obtain a numerically treatable model in which we can study the cosmological implications of the ET, our model will be based on just two degrees of freedom, the inflaton $\varphi$ and a scalar field $\chi$, representing the energy absorbing boundary.

The field $\chi$ is used to model in field theory a variety of
boundary barotropic fluids by endowing $\chi$ with a suitable
exponential potential, as we now explain. Since we do not know the
equation of state of the interacting tensionless strings which are
being produced in the small instanton transitions and into which the
energy of the inflaton is fed, one would like to adopt a general
barotropic perfect fluid with equation of state \beqa P=w \rho \; ,
\eeqa which will absorb part of the inflaton's energy. The influence
of the ET on the cosmological perturbations can then be studied for
various values of the parameter $w$. This way one could e.g.~model
radiation, matter or cosmic strings as possible effective components
being generated on the brane in the collision. Now in a field theory
framework one should rather describe the perfect fluid by a suitable
field. In fact, it had been shown in \cite{Lucchin:1984yf} that a
scalar field $\chi$ with an exponential potential \beqa
\label{chi-potential-1} V(\chi) \sim
e^{-\sqrt{\frac{2}{q}}\frac{\chi}{M_{\rm P}}} \eeqa leads to a
power-law evolution of the scale-factor, $a(t)\sim t^q$, even when
$\frac{1}{3}\leq q \leq 1$. This type of evolution would also result
from a perfect fluid with equation of state parameter \beqa
w=\frac{2}{3q}-1 \; . \eeqa Hence we will model the perfect fluid in
a field theory description by the scalar $\chi$ together with the
above exponential potential. The reader should bear in mind that the
energy content of the fluids are subdominant with respect to the
inflaton's energy and thus the transferred energy does not dominate
the evolution of the background. Our work thus differs from others,
such as \cite{Burgess:2005sb}, in that respect.

Our potential for the two scalar field model
\beqa
V(\varphi,\chi) =
V(\varphi) + W(\chi,\varphi)
\eeqa
consists therefore of two
different components. First, there is the inflaton potential \beqa
V(\varphi)&=&v(\varphi) \exp\left(-\sqrt{\frac{2}{p(\varphi)}}
\frac{(\varphi-\varphi_s)}{M_{\rm P}}\right) \; ,
\eeqa
with $\varphi_s$ being the value of the inflaton at the step (brane
collision). This type of potential results from an M5-brane
collision in cascade inflation, see eq.~\eqref{Potential}. The
amplitude $V_N$ and parameter $p_N$ vary with $N$ in a collision. We
model this step-like variation using an inflaton dependent amplitude
and parameter (the index $i$ refers to the initial state before
hitting the step, whereas the index $f$ refers to the final state
after hitting the step)
\begin{eqnarray}\label{vwpq}
v(\varphi)&=&\frac{U_f+U_i}{2}+\frac{U_f-U_i}{2}\tanh\left(\frac{\varphi-\varphi_s}{\Delta \varphi}\right)
\\
p(\varphi)&=&\frac{p_f+p_i}{2}+\frac{p_f-p_i}{2}\tanh\left(\frac{\varphi-\varphi_s}{\Delta \varphi}\right)
\end{eqnarray}
with a tanh dependence, which interpolates between $-1$ and $1$ and smoothes the step, while the parameter $\Delta \varphi$ governs the smoothed step's width. As said earlier, we will, for simplicity, focus our attention on a single step (single brane collision) in the inflaton's potential.

Second, there is the scalar $\chi$ which has to absorb a certain amount of energy in the course of the collision, as can be seen as follows. Before the collision, there exists only the inflaton, $\varphi$, with an exponential potential with
parameter $p_i$ and amplitude $U_i$. During inflation the inflaton's value increases until it approaches the step at a
value $\varphi_s$. Here the exponent drops to $p_f$ and the potential's amplitude to $U_f$. The difference between initial and final inflaton potential energy has to be absorbed by the boundary fluid, i.e.~is transferred to the field, $\chi$. The potential for $\chi$ should be of the form given in eq.~\eqref{chi-potential-1} for the reasons explained above. However, since energy is transferred to $\chi$, the amplitude of its potential will change in time and thus depend on $\varphi$. Similarly, we can expect the nature of the fluid to change during the collision as e.g.~a tensionless string component is created which hadn't been there before the collision. Therefore, we model the potential for $\chi$ by
\begin{eqnarray}\label{W}
W(\chi,\varphi)&=&w(\varphi) \exp\left(-\sqrt{\frac{2}{r(\varphi)}}
\frac{\chi}{M_{\rm P}}\right),
\end{eqnarray}

with inflaton dependent amplitude $w(\varphi)$ and parameter $r(\varphi)$. These are again expressed in terms of tanh functions
\begin{eqnarray}\label{wr}
w(\varphi)&=&\frac{U_i-U_f}{2}\left(1+\tanh\left(\frac{\varphi-\varphi_s}{\Delta \varphi}\right)\right)\\
r(\varphi)&=&\frac{q+p_i}{2}+\frac{q-p_i}{2}\tanh\left(\frac{\varphi-\varphi_s}{\Delta
\varphi}\right)
\end{eqnarray}
to describe a smoothed out step. In this way we arrive at a coupling between $\varphi$ and $\chi$.

Before encountering the step, the scalar $\chi$ describes a fluid with equation of state parameter $w_i = \frac{2}{3p_i} - 1$. The potential of $\chi$ is almost zero because its amplitude is almost vanishing. This changes quickly during the collision, at which the amplitude raises to $U_i-U_f$ (the ET coming from the inflaton) and the equation of state parameter soon adjusts itself at $w = \frac{2}{3q} - 1$. In what follows, we will mostly focus on the case where the inflaton's energy is transferred to an effective
radiation component ($q=1/2$) but will also consider matter ($q=2/3$) or a network of cosmic strings component ($q=1$). We concentrate on the radiation case first and choose for definiteness the following parameter values
\begin{equation}\label{par1}
p_i=40.138\; , \qquad p_f=36.598\; , \qquad \frac{(U_i-U_f)}{U_i}=0.068 \; ,
\end{equation}
\begin{equation}\label{par2}
\varphi_s=1.477 M_{\rm P}\; , \qquad \Delta \varphi=10^{-3} M_{\rm P} \; .
\end{equation}
\begin{figure}[t]
\includegraphics[angle=0, width=80mm, height=70mm]{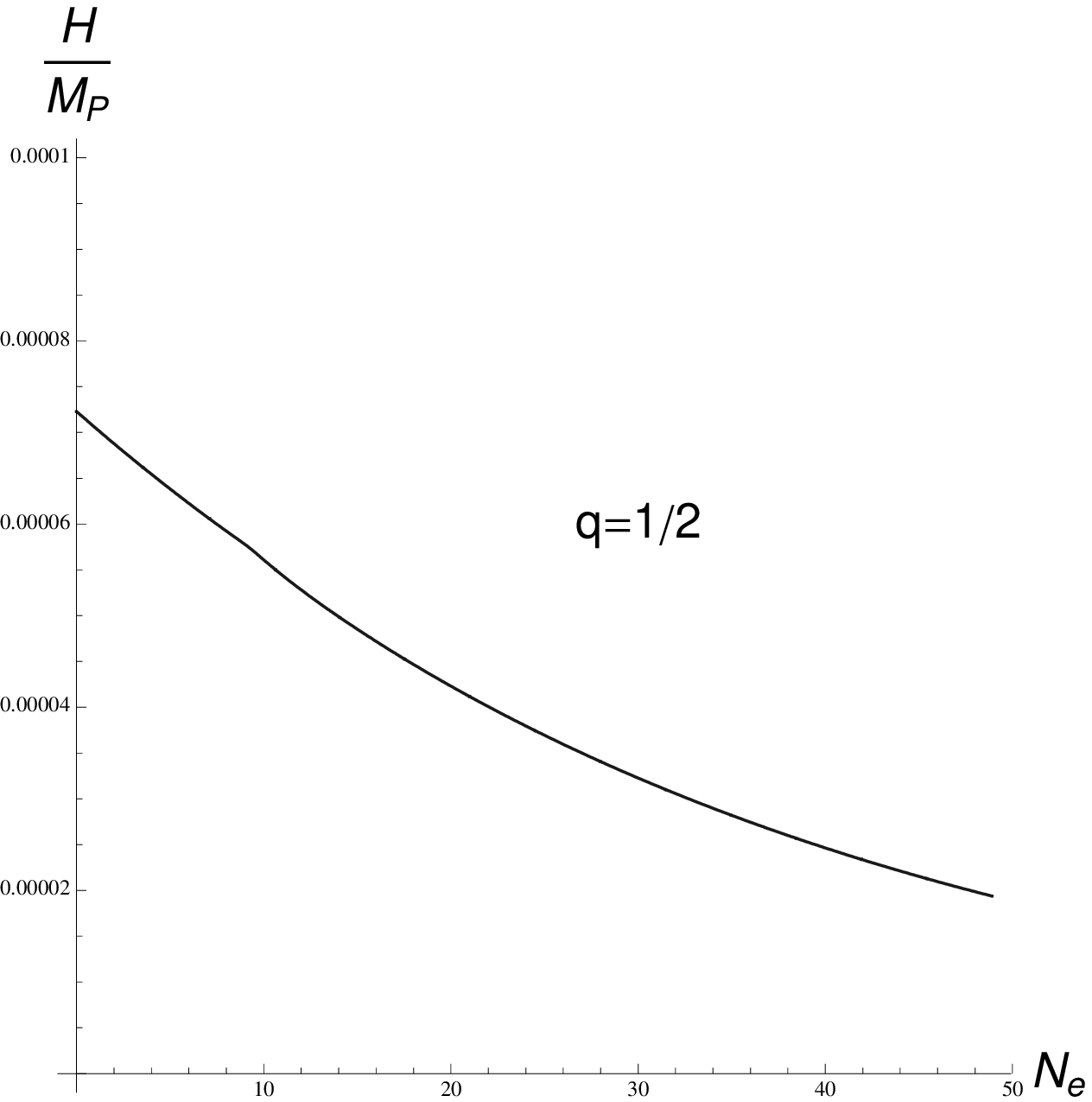}
\includegraphics[angle=0,width=80mm, height=70mm]{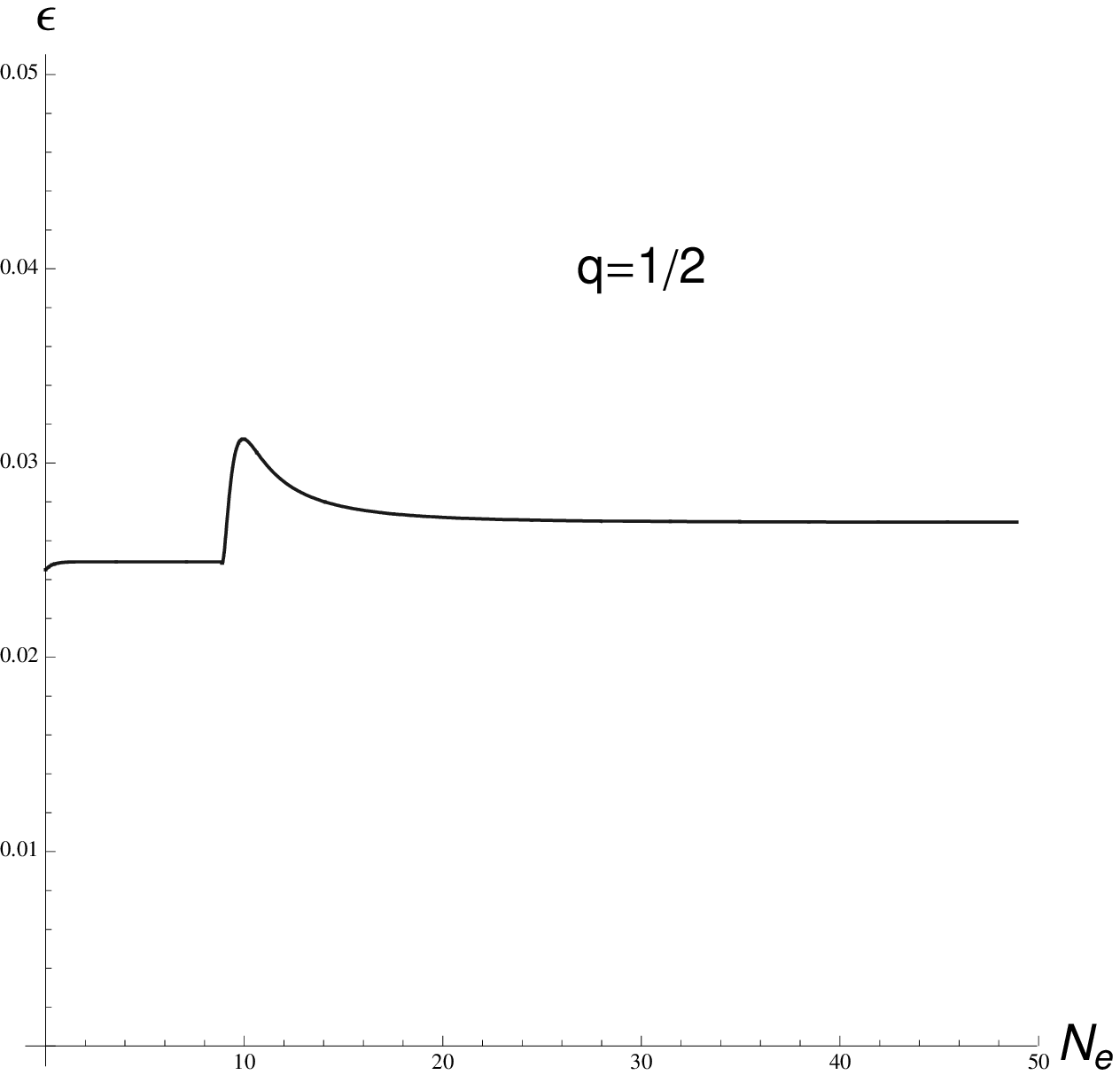}
\caption{The graphs show the evolution of the Hubble parameter and
the first slow-roll parameter as a function of the number of
e-foldings. Around $N_e\simeq 10$ the inflaton's potential energy
$(U_i-U_f)$ is transferred to the $\chi$
field.}\label{Hubble-epsilon}
\end{figure}
\begin{figure}[t]
\includegraphics[angle=0, scale=.72]{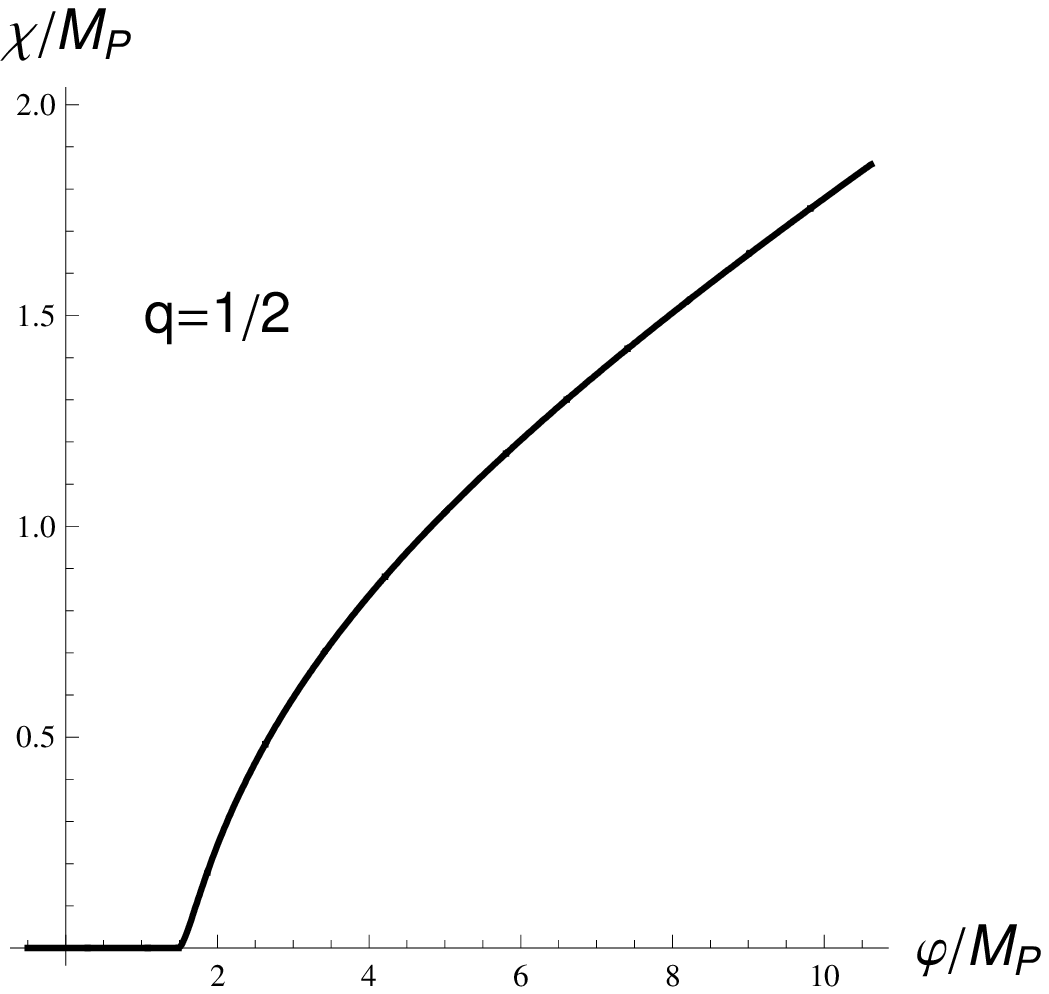}
\includegraphics[angle=0, width=75mm, height=70mm]{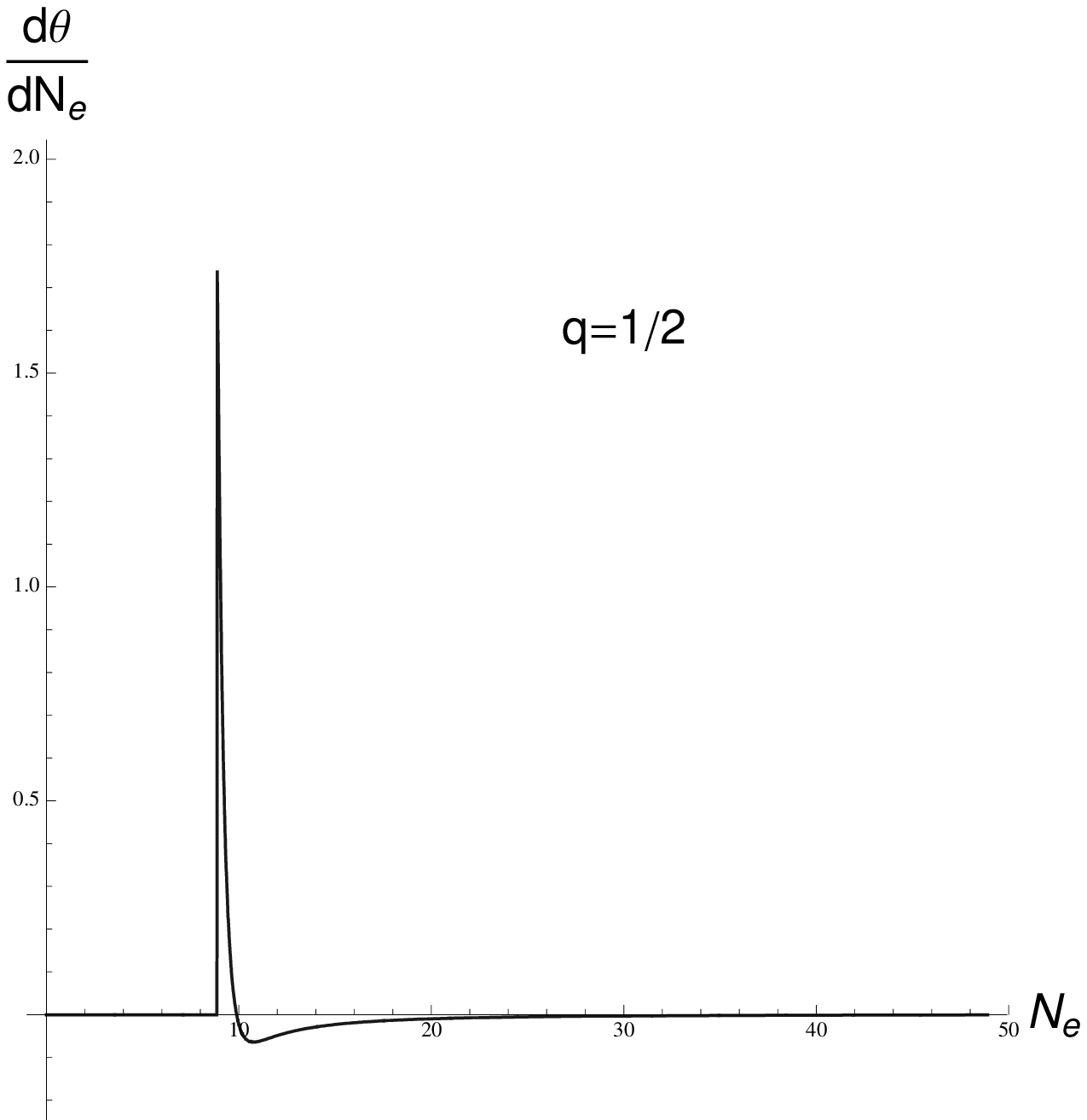}
\caption{The left graph shows the evolution of the fields $\varphi$
and $\chi$. They move upwards from left to right along the black trajectory, whose very first part coincides with the $\varphi$ axis. The right graph displays $d\theta/dN_e$ as a function of the number of e-foldings $N_e$. A sharp turn in field space at the collision is clearly visible.}\label{field-evol-dtheta-dN}
\end{figure}
The initial value of $\chi$ is chosen to be zero, $\chi_{i}=0$, such
that the inflaton's potential energy $U_i-U_f$ is transferred to the
$\chi$ field.

\subsection{Cosmological Evolution}

Fig.~\ref{Hubble-epsilon} shows the evolution of the Hubble
parameter and the first slow-roll parameter of the model as a
function of the number of e-foldings. Around $N_e\simeq 10$ the step
is encountered (brane collision takes place) and the fraction
$(U_i-U_f)/U_i$ of the inflaton's potential energy is transferred to
the $\chi$ field. Since the $\chi$ direction in the total potential
$V(\varphi,\chi)$ is much steeper than the inflaton direction, its
energy content redshifts within a few e-foldings and the background
evolves solely under the influence of the final inflaton potential
\beqa V(\varphi) \rightarrow U_f
\exp\left(-\sqrt{\frac{2}{p_f}}\frac{(\varphi-\varphi_s)}{M_{\rm
P}}\right) \eeqa after that. As the slope of the potential increases
after the energy of $\chi$ redshifts, the slow-roll parameter,
$\epsilon$, settles to a slightly larger value, see the right graph
in fig.~\ref{Hubble-epsilon}.

The trajectory of $\varphi$ and $\chi$ in field space and the
evolution of their combination $\theta$, which had been defined in
eq.~\eqref{theta}, is graphed in fig.~\ref{field-evol-dtheta-dN}.
There is a sharp turn in the classical trajectory of the fields,
when the inflaton, $\varphi$, transfers part of its energy to
$\chi$. As we will see, with this bent in the trajectory, the
curvature perturbations are strongly fed by entropy perturbations.
\begin{figure}[t]
\includegraphics[angle=0, width=80mm, height=70mm]{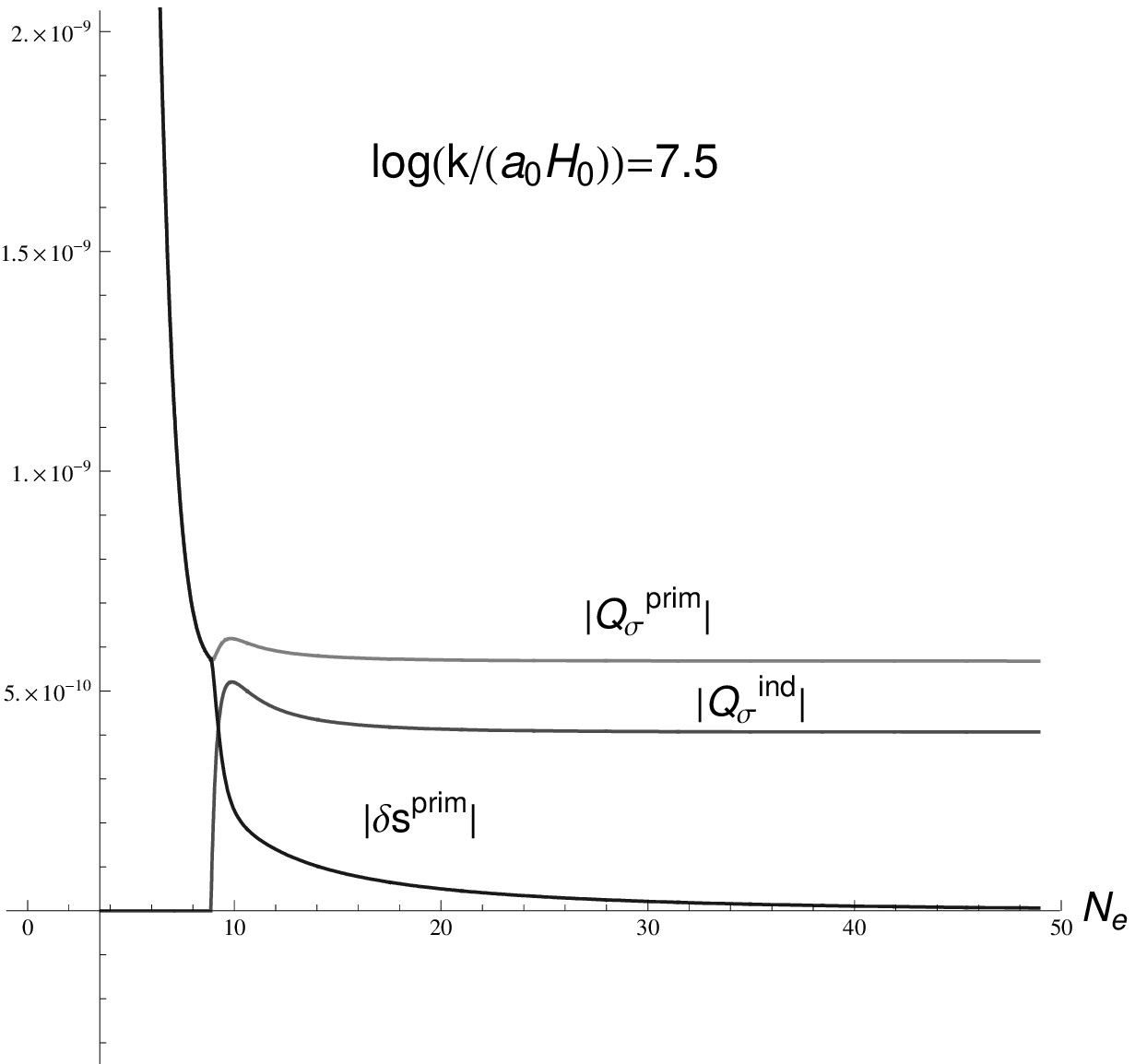}
\includegraphics[angle=0, width=80mm, height=70mm]{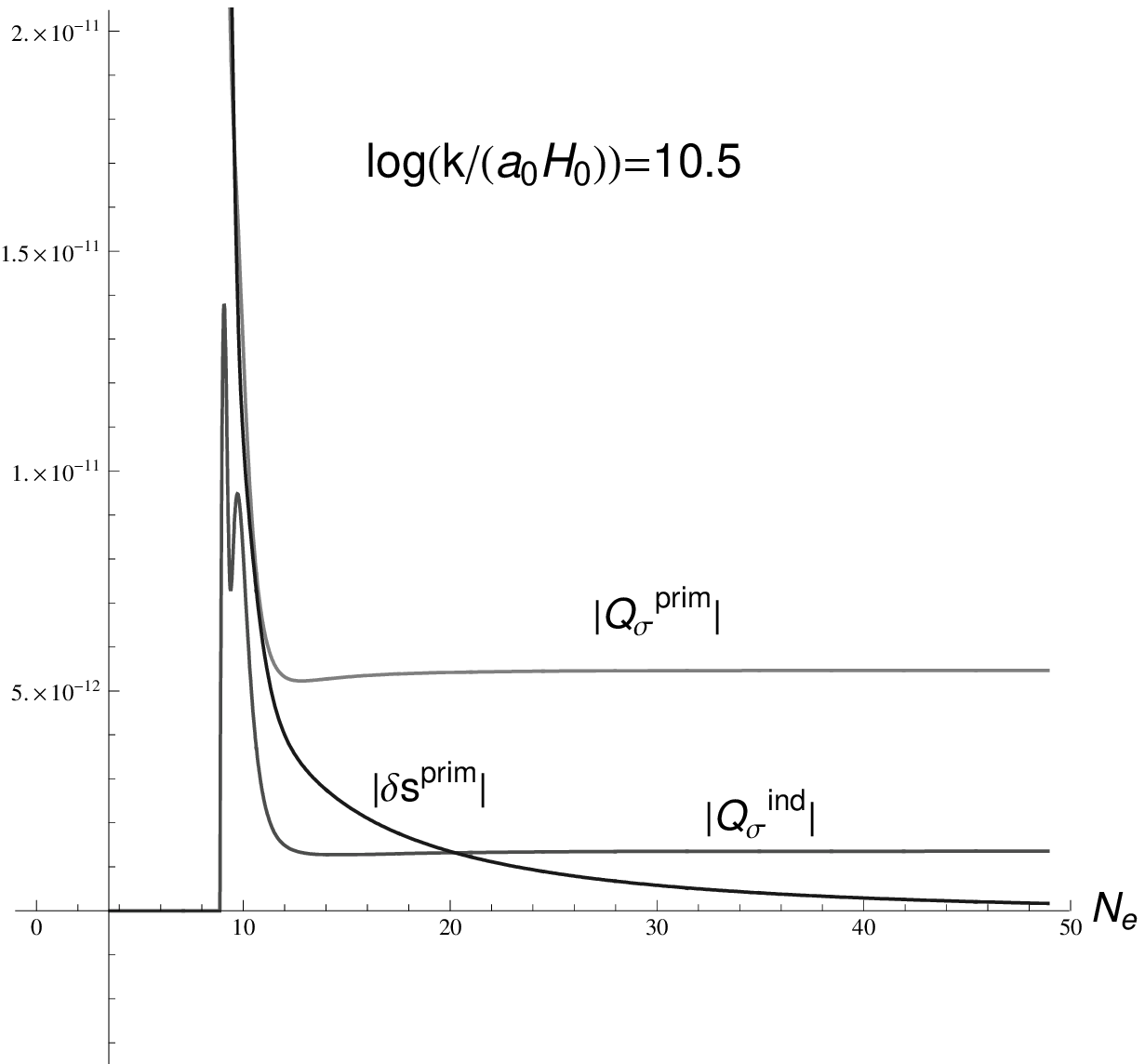}
\caption{The left and right graphs show the evolution of
$|Q_{\sigma}^{\rm prim}|$, $|Q_{\sigma}^{\rm ind}|$ and $|\delta
s^{\rm prim}|$} as functions of $N_e$ for two comoving wave-numbers that exit the
horizon before respectively after the ET. \label{single-mode}
\end{figure}
We will now calculate the curvature and isocurvature spectra for our
energy exchanging two field model.

\begin{figure}[t]
\includegraphics[angle=0, width=80mm, height=70mm]{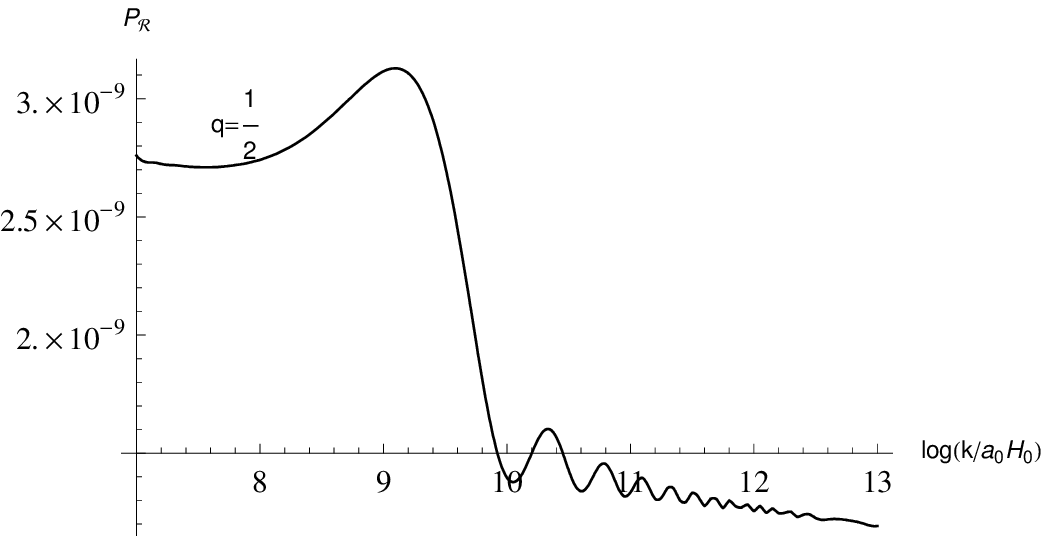}
\includegraphics[angle=0, width=80mm, height=70mm]{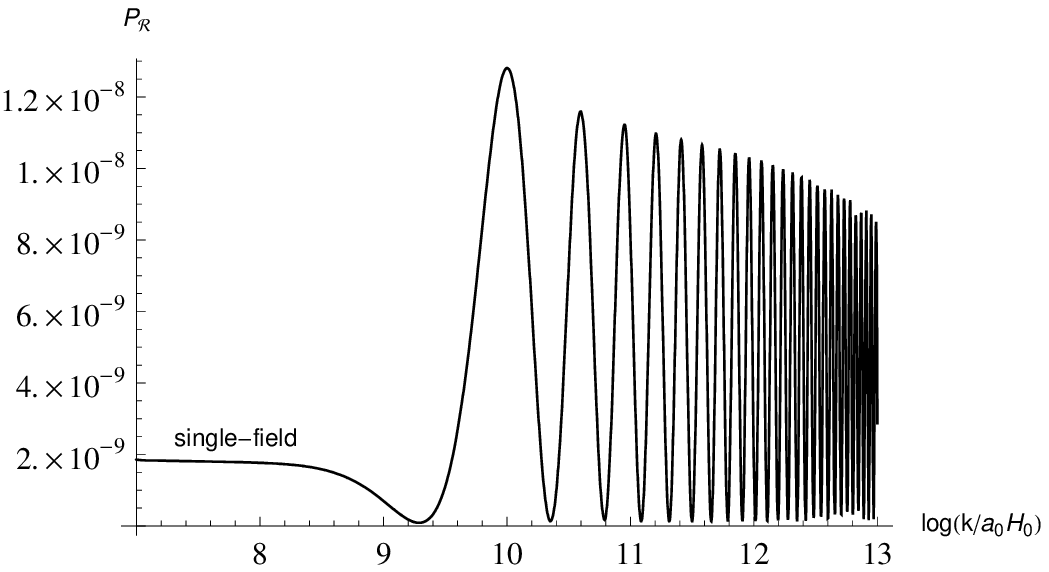}
\caption{The left graph shows the adiabatic spectrum vs.
$\log(k/a_0H_0)$ for the modes that exit the horizon around the
decay time. It is assumed that the energy of the decay products
redshifts as radiation, $U_f/U_i=0.932$ and $\Delta \phi=10^{-3}
M_{\rm P}$. The right graph shows the adiabatic power spectrum for
the single field case with an inflaton potential having a step of
equal height. The oscillations in the single field case last much
longer than in the two field case with
ET.}\label{adiabatic-spectrum-p=0.5-singlefield}
\end{figure}

\section{Adiabatic and Isocurvature Perturbations} \label{chapter-Spectra}

For the two-field model described above, we integrate the equations
of motion for the curvature and the isocurvature perturbations ,
eqs. (\ref{perturbations}) and (\ref{perturbationsDelSig}), whose
derivation is outlined in Appendix.

\subsection{Evolution of Curvature and Entropy Perturbations.}

Fig.~\ref{single-mode} shows the result for the evolution of
curvature and entropy fluctuations for two modes: one that exits the
horizon before the ET at $\log(k/a_0H_0)=7.5$, and a second mode
which exits it after the ET at $\log(k/a_0H_0)=10.5$. The subscript
$0$ denotes the values of the scale factor and Hubble parameter at
the beginning of inflation. Since we have to deal with two
independent physical degrees of freedom, we perform the integration
twice to account for the two independent quantum fluctuations. In
the first run, we assume that $Q_\sigma$ is initially in the
Bunch-Davies vacuum and that $\delta s$ initially vanishes. In this
manner, we obtain the primordial curvature perturbations
$Q_\sigma^\mathrm{prim}$ and the induced isocurvature perturbations
$\delta s^\mathrm{ind}$. In the second run we interchange the
initial conditions for $Q_\sigma$ and $\delta s$ to obtain
$Q_\sigma^\mathrm{ind}$ and $\delta s^\mathrm{prim}$. The total
amplitude of the curvature perturbations is given by:
\begin{equation}\label{Qtot}
{|Q_{\sigma}^{\rm tot}|}^2={|Q_{\sigma}^{\rm
ind}|}^2+{|Q_{\sigma}^{\rm prim}|}^2
\end{equation}
and the amplitude of the total isocurvature perturbations is given
by an analogical expression. In the single-field case, we perform
just one integration.

Deep inside the Hubble radius, the two perturbations evolve
independently in the same way, up to a slow overall rotation which
practically does not change the amplitudes or the correlations.
After the Hubble radius crossing $|Q_{\sigma}^{\rm prim}|$
approaches the value it would have obtained in absence of any
isocurvature perturbations. The induced perturbations,
$Q_{\sigma}^{\rm ind}$ and $\delta s^{\rm ind}$ are practically
negligible inside the Hubble radius. The former can be generated on
the super-Hubble scales, where its EOM in the slow-roll
approximation reads \cite{Starobinsky:1994mh,Gordon:2000hv}:
\begin{equation}
\frac{1}{H} \dot{\mathcal{R}} = \frac{k^2}{\dot{H}a^2} \Phi
-\frac{2}{H}\dot{\theta}\mathcal{S} \, ,
\end{equation}
given that the background trajectory in the field space is
sufficiently curved. Also $|Q_{\sigma}^{\rm ind}|$ for such modes is
much smaller than $Q_{\sigma}^{\rm prim}$ and consequently we do not
see considerable enhancement at such scales.

 The behavior of the field valued function $d\theta/dN$ is displayed in the right graph of
fig.~\ref{field-evol-dtheta-dN}. Before the ET, this function is
zero and thus no induced curvature perturbations, $|Q_{\sigma}^{\rm
ind}|$, are generated. As the ET from $\varphi$ to $\chi$ happens,
the sharp turn in the classical trajectory creates a spike in
$d\theta/dN$. This will lead to a considerable increment in the
curvature perturbations on super-Hubble scales, due to interaction
with isocurvature perturbations. As the energy of the $\chi$-field
redshifts, $d\theta/dN$ becomes zero again and the amplitude
$|Q_{\sigma}^{\rm ind}|$ becomes frozen. This results in an overall
enhancement of the amplitude of the curvature modes that exit the
horizon before the ET. For modes that exit the horizon after the ET,
the function $|Q_{\sigma}^{\rm ind}|$ undergoes some modulated
oscillations before becoming constant at super-horizon scales. As we
will see, these oscillations will imprint themselves as modulated
oscillations on the curvature spectrum.

\subsection{Curvature and Entropy Spectra}

The left plot of fig.~\ref{adiabatic-spectrum-p=0.5-singlefield}
demonstrates the dependence of the adiabatic spectrum on $\log(k/a_0
H_0)$. For comparison, we have also shown the power spectrum for a
single inflaton model with exponential potential possessing a step
of same height. This model is obtained by setting the potential for
the $\chi$ field, $W(\chi,\varphi)$, and therefore also the coupling
to $\varphi$, to zero. The height of the potential step in the
single field case is entirely transferred into kinetic energy of the
inflaton \cite{Adams:2001vc}. One point that easily gets noticed by
comparing these two graphs is that the modulated oscillations in the
power spectrum for the two field case decay much faster than in the
single field case. For $U_f/U_i=0.932$, the modulated oscillations
last for four decades in $k$ for the two field case, whereas in the
single field case they continue for more than eight decades. Also
the amplitude of those modes that exit the horizon before the decay
is increased by $41\%$. As mentioned above, this is due to the
strong interaction of the adiabatic and isocurvature perturbations,
when part of the inflaton's energy is transferred to $\chi$.
\begin{figure}[t]
\includegraphics[angle=0, width=80mm, height=70mm]{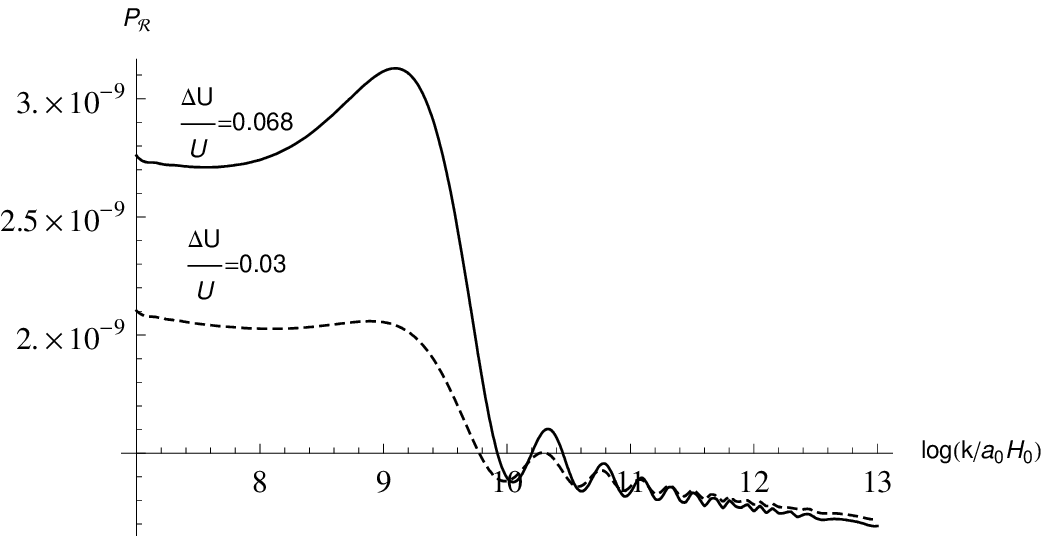}
\includegraphics[angle=0, width=80mm, height=70mm]{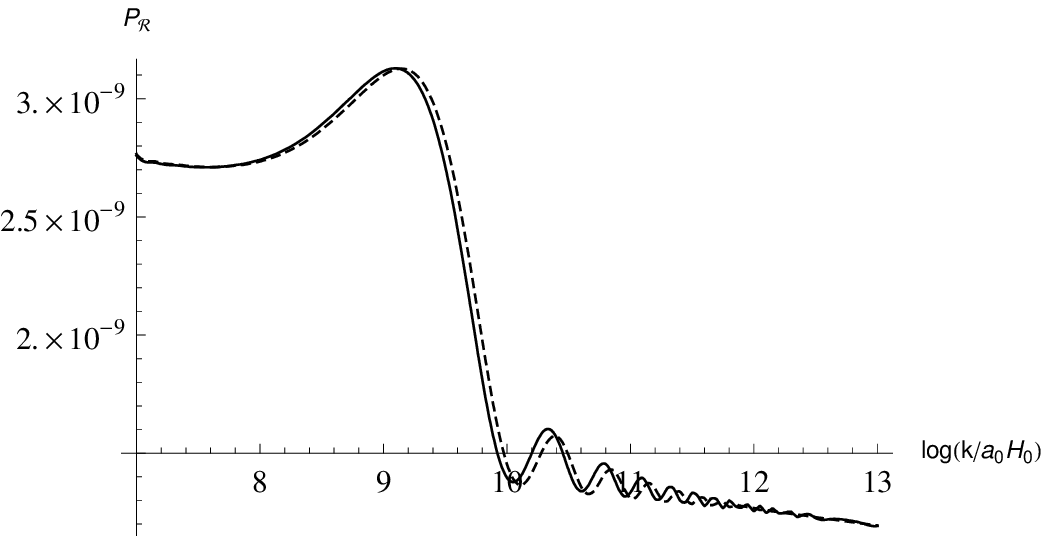}
\caption{The left graph shows the curvature spectra vs.
$\log(k/a_0H_0)$ for $\Delta U/U_i=0.068$ and $\Delta U/U_i=0.03$.
The right graph shows the adiabatic power spectra vs.
$\log(k/a_0H_0)$ for $\Delta\varphi=10^{-2} M_{\rm P}$ (black solid
line) and $\Delta\varphi=10^{-3} M_{\rm P}$ (dashed grey line)
}\label{DeltaU-Deltaphi}
\end{figure}

\begin{figure}[t]
\includegraphics[angle=0, width=80mm, height=70mm]{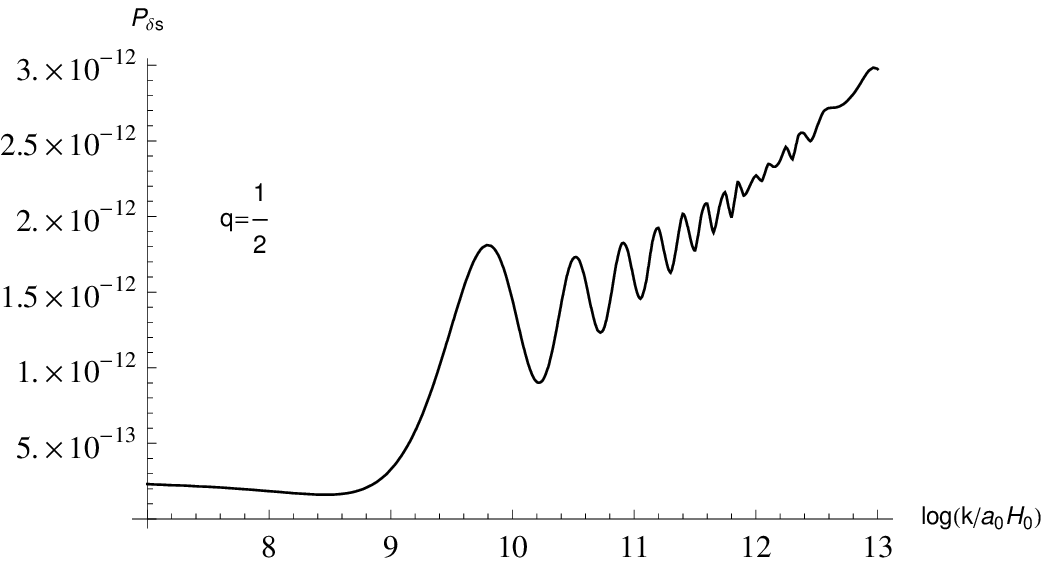}
\includegraphics[angle=0, width=80mm, height=70mm]{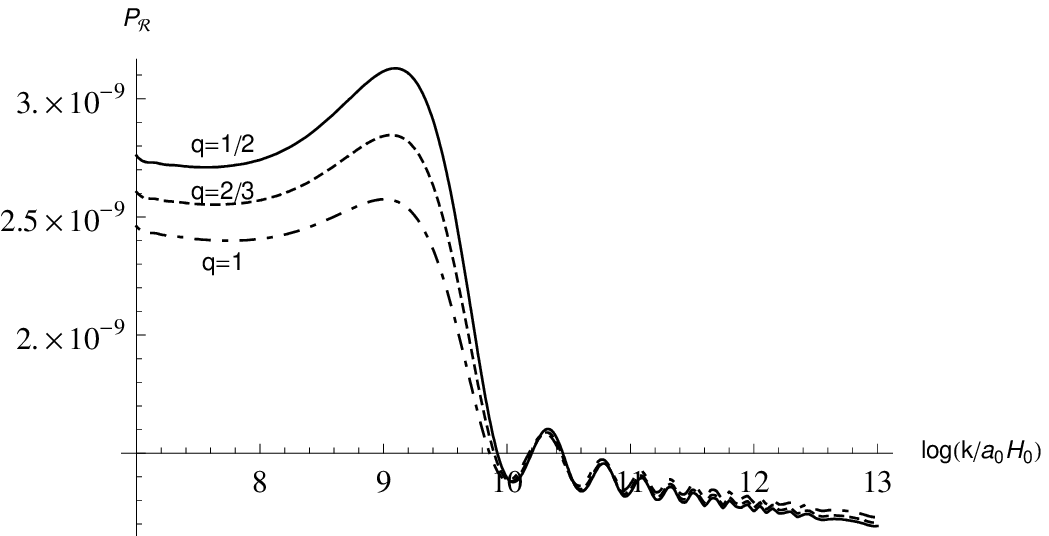}
\caption{The left graph shows the entropy spectrum vs.
$\log(k/a_0H_0)$ for $\Delta U/U_i=0.932$ and $\Delta
\varphi=10^{-3}M_{\rm P}$. The right graph shows the dependence of
the adiabatic spectrum on $q$.} \label{entropy-spectrum-q-compared}
\end{figure}

In the slow roll approximation, the EOM of the isocurvature
perturbations on the super-Hubble scales can be approximated by:
\begin{equation}\label{solution-entropy}
\frac{\mathrm{d} \delta s}{\mathrm{d}N_e}\simeq -\eta_{ss} \delta
s\, .
\end{equation}
Since the potential is typically much more curved in the direction
orthogonal to the trajectory in the field space than in the
direction along the trajectory. Hence, the isocurvature
perturbations decay exponentially from their corresponding value at
the horizon crossing. The shape of the isocurvature spectrum is
given in fig.~\ref{entropy-spectrum-q-compared}. Their corresponding
value is sub-dominant with respect to the curvature perturbations by
a factor which varies between $10^{-3}$ to $10^{-4}$ at different
scales. One can see a slight amplification followed by modulated
oscillation at the scales that leave the horizon around the ET.

We now consider the effect of changing \beqa \Delta U/U \equiv
(U_i-U_f)/U_i \; , \eeqa keeping all other parameters constant. The
left graph of fig.~\ref{DeltaU-Deltaphi} shows the adiabatic spectra
for $\Delta U/U=0.068$ and $\Delta U/U=0.03$. Reducing the energy
which gets transferred to the $\chi$ field by a factor of $\sim
2.25$, decreases the amount of increment in the adiabatic
perturbations at wave numbers smaller than $k_{{\rm ET}}\equiv
aH|_{\rm ET}$ by a factor of $\sim 4.2$. The amplitude of the
modulated fluctuations at $k>k_{{\rm ET}}$ decreases too, even
though the frequency of the oscillations remains more or less the
same.

We also considered the effect of changing the parameter $\Delta
\varphi$ which physically corresponds to the decay width of the
inflaton. The result is displayed in the right graph of
fig.~\ref{DeltaU-Deltaphi}. The smaller this parameter is, the
faster the energy from inflaton, $\varphi$, transfers to $\chi$. By
decreasing this parameter, the adiabatic power spectrum is shifted
slightly toward larger scales. This is intuitively understandable,
as in this case the inflaton energy, $\Delta U$, is exchanged faster
and thus the resulting oscillations start at larger scales.

So far we have assumed that the decay product $\chi$ has a potential
which makes it redshift like radiation, $q=1/2$. We now relax this
constraint and consider how the perturbations will evolve if we
choose $q$ appropriate for matter, $q=2/3$, or a web of cosmic
strings, $q=1$. Fig.(\ref{entropy-spectrum-q-compared}) shows how
the curvature spectrum changes once $\chi$ decays like radiation.
For a fixed amount of energy transferred to $\chi$ field, increasing
$q$, reduces the amount of amplification of curvature spectrum at
wave-numbers $k<k_{\rm ET}$. Thus the least amount of amplification
at such scales occurs for $q=1$. The amplitude and frequency of the
oscillations are more or less independent from parameter $q$.
\begin{figure}[t]
\includegraphics[angle=0, width=85mm, height=70mm]{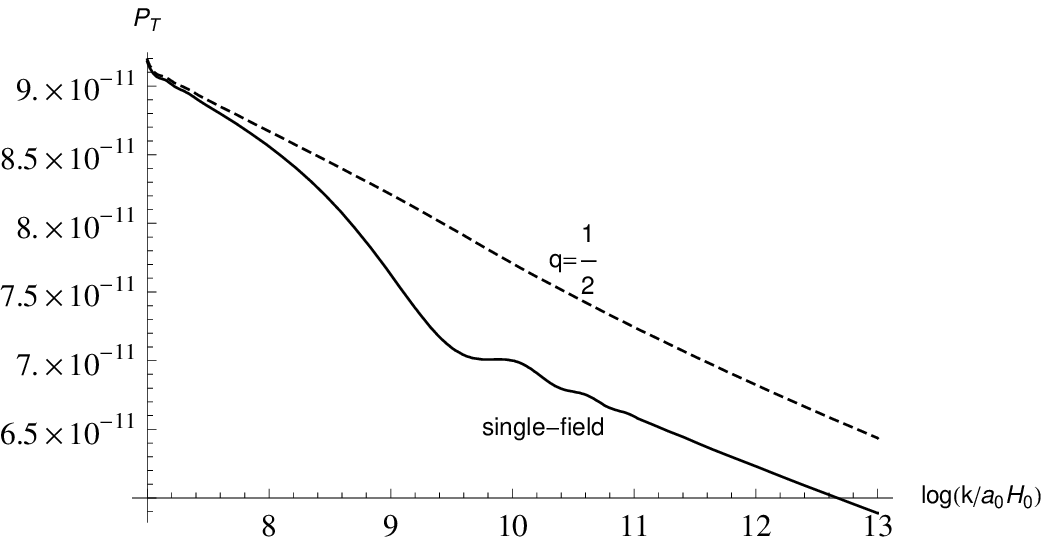}
\includegraphics[angle=0, width=70mm, height=70mm]{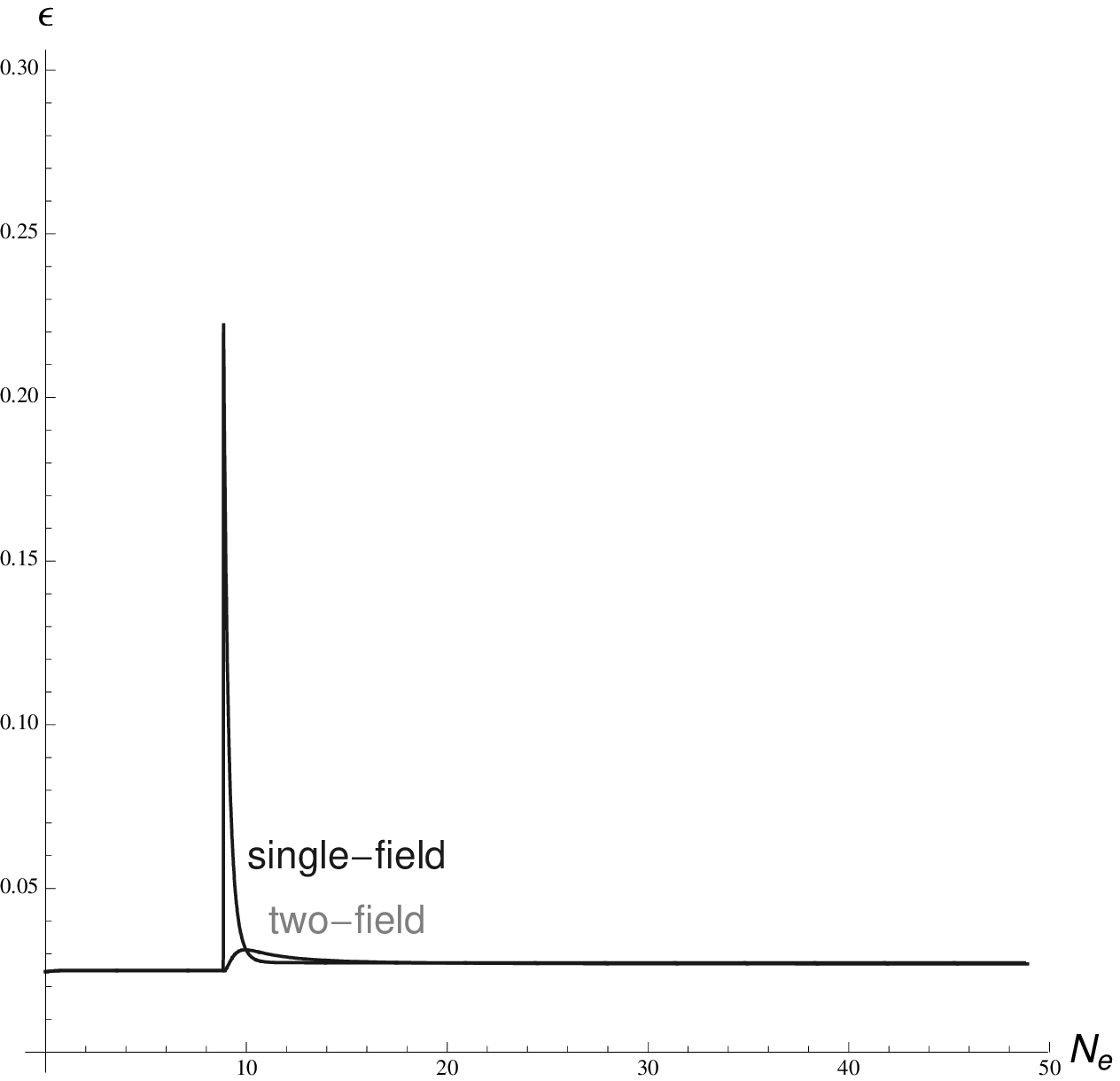}
\caption{The left graph shows the tensor spectra vs. $\log(k/a_0H_0)$
for $\Delta U/U_i=0.932$. Solid and dashed lines respectively
represent the single field and two field cases. The right plot shows the
dependence of the slow-roll parameter $\epsilon$ on $N_e$.} \label{tensor-spectra}
\end{figure}

\subsection{Tensor Spectrum}

Finally, we are investigating the tensor spectra of the energy exchanging two field inflation model. The left plot of fig.~\ref{tensor-spectra} shows the profile of gravity
waves for the modes that exit the horizon during the ET. For
comparison we have also plotted the tensor spectrum for the single field case, in which the energy of the step in the $\varphi$ potential is snatched by the kinetic energy of $\varphi$ itself. In the single field case the spectrum displays some oscillations for the modes
that exit the horizon around the ET. This could be understood by
comparing the variations of $\epsilon$ for the two cases. Tensor
perturbations satisfy the following equation \cite{Lidsey:1995np}
\begin{equation}\label{tensor-perturbations}
p''_{k}+\left(k^2-\frac{a''}{a}\right)p_k=0.
\end{equation}
The quotient $a''/a$ can be written in terms of the slow-roll parameter $\epsilon$ as
\begin{equation}\label{apprima}
\frac{a''}{a}=2a^2 H^2(2-\epsilon) \; .
\end{equation}
To understand its implication for the tensor spectrum, we exhibit in the right plot of fig.~\ref{tensor-spectra} the evolution of $\epsilon$ with $N_e$ for the two field
and single field cases. In both cases, during the non-slow-roll
phases, there are sharp spikes in $\epsilon$. However, for the
single field case, the spike in $\epsilon$ is much greater than in the
two field case. This large variation in $\epsilon$ in the
single field case leads to modulated wiggles on the amplitude of
its tensor spectrum and explains the difference between the single and two field cases seen in the left part of fig.~\ref{tensor-spectra}.

This result for the single field case should be
contrasted with the results of \cite{Adams:2001vc}, where no observable signature in the tensor power spectrum had been observed. However, one should note
that the energy difference in the step was chosen much smaller in
\cite{Adams:2001vc} and thus $\epsilon$ would not exhibit such a
sharp spike.

Furthermore, it can be seen from the left plot in
fig.~\ref{tensor-spectra} that the amplitude of the tensor spectrum
decays much faster in the single field than in the two field case.
This is because the liberated energy from the potential's step
transforms in the single field case to the kinetic energy of the
inflaton $\varphi$, which redshifts like $a^{-6}$. On the contrary,
in the two field case the released energy is absorbed by the $\chi$
field which redshifts much slower. This will cause the the Hubble
parameter in the single field case to diminish much faster. As the
amplitude of the tensor perturbation is roughly given by $H/2\pi$,
this explains the smallness of the tensor spectra in the single
field case compared to the larger tensor spectra of the two field
case which we see in the left plot of fig.~\ref{tensor-spectra}.

\section*{Acknowledgments}

A.A.~is partially supported by the Natural Sciences and Engineering
Research Council of Canada. A.K.~is supported by the German Research
Foundation (DFG) and the Transregional Collaborative Research Centre
TR33 ``The Dark Universe''. K.T. is partially supported by the grant
MNiSW N202 176 31/3844 and by TOK Project MTKD-CT-2005-029466. K.T.
also acknowledges support from the Foundation for Polish Science
through its programme HOMING.

\appendix

\section*{Appendix: Curvature and Isocurvature Perturbations in Two-field
Inflation}

The calculation of cosmological perturbations in the multi-field
inflation is an extensively studied topic. Nonetheless, we would
like to review the basic notation, results and, in particular, the
equations of motion for the perturbations that we solve numerically.
In this section, we shall follow closely the presentation of
\cite{Lalak:2007vi}.

A two-scalar-field system coupled to gravity is described by an
action of the form
\begin{equation}\label{action}
S=\int \mathrm{d}^4 x \sqrt{-g}\left(\frac{M_{\rm
P}^2}{2}R-\frac{1}{2}\partial_{\mu}\varphi\partial^{\mu}
\varphi-\frac{1}{2}\partial_{\mu}\chi\partial^{\mu}
\chi-V(\varphi,\chi)\right) \; ,
\end{equation}
where $M_{\rm P}$ is the reduced Planck mass, $M_{\rm P}\equiv (8\pi
G)^{-1/2}$.
The homogeneous and isotropic FRW background with metric
\begin{equation}\label{metric}
ds^2=-dt^2+a(\mathrm{t})^2 d{\mathbf{x}}^2,
\end{equation}
is governed by the equations of motion (EOM) for the two scalar
fields:
\begin{eqnarray}\label{eom-phi}
\ddot{\varphi}+3H\dot{\varphi}+V_{,\varphi}&=&0\\
\label{eom-chi} \ddot{\chi}+3H\dot{\chi}+V_{,\chi}&=&0 \; .
\end{eqnarray}
Subscripts $\varphi$ and $\chi$ denote partial derivatives with
respect to the corresponding field and a dot denotes a derivative
with respect to the cosmic time, $t$. The gravitational background
evolves according to Friedmann-Lem\^aitre equations:
\begin{eqnarray}
H^2 &=&\frac{1}{3M_{\rm P}^2}\left(\frac{1}{2}\dot{\varphi}^2+\frac{1}{2}\dot{\chi}^2+V(\varphi,\chi)\right)  \\
  \dot{H} &=& -\frac{1}{2M_{\rm P}}\left(\dot{\varphi}^2+\dot{\chi}^2\right) \; ,
\end{eqnarray}
where $H$ is the Hubble parameter, defined as $H \equiv
\frac{\dot{a}}{a}$.

To study the linear perturbations for this theory, we start with the
longitudinal gauge for the metric \cite{Mukhanov:1990me}. In the
absence of any anisotropic stress-energy tensor the scalar
perturbation of the gravitational background reads:
\begin{equation}\label{metric-long}
ds^2=-\big(1+2\Phi(\mathrm{t},\mathbf{x})\big)d\mathrm{t}^2 +
a(\mathrm{t})^2 \big(1-2\Phi(\mathrm{t},\mathbf{x})\big)
d\mathbf{x}^2.
\end{equation}
The scalar fields are also perturbed around  their homogeneous
parts,
\begin{equation}\label{fields-perturb}
\varphi(\mathrm{t},\mathbf{x})=\varphi(\mathrm{t})+\delta
\varphi(\mathrm{t},\mathbf{x})~~~~~{\rm
and}~~~~~\chi(\mathrm{t},\mathbf{x})=\chi(\mathrm{t})+\delta
\chi(\mathrm{t},\mathbf{x}),
\end{equation}
These perturbations introduce an $\mathbf{x}$  dependence which was
not present in the homogeneous and isotropic gravitational and
scalar field backgrounds. To determine the perturbations, one
therefore has to insert the perturbed metric and scalar fields into
the full Einstein field equations and/or Bianchi identities, and the
full scalar field EOMs.

Since the perturbations of the metric and the scalar fields are not
independent, it is useful to introduce gauge-invariant
Mukhanov-Sasaki variables
\begin{equation}\label{Qphi-Qchi}
Q_{\varphi}\equiv \delta\varphi+\frac{\dot{\varphi}}{H}\Phi~~~~~{\rm
and}~~~~~Q_{\chi}\equiv \delta\chi+\frac{\dot{\chi}}{H}\Phi \; .
\end{equation}
They represent the scalar field fluctuations in the flat gauge. It
follows from the EOMs that their Fourier-components obey the coupled
differential equations\footnote{Even though, we will work with
Fourier components hereafter, we will not show explicitly the
subscript $\mathbf{k}$ which would denote the fluctuation with
comoving wave-number $\mathbf{k}$.}
\begin{eqnarray}\label{eqs}
\ddot{Q}_{\varphi}+3H\dot{Q}_{\varphi}+\left(\frac{k^2}{a^2}+C_{\varphi\varphi}\right)Q_{\varphi}+C_{\varphi\chi}Q_{\chi}&=&0\\
\ddot{Q}_{\chi}+3H\dot{Q}_{\chi}+\left(\frac{k^2}{a^2}+C_{\chi\chi}\right)Q_{\chi}+C_{\chi\varphi}Q_{\varphi}&=&0
\; ,
\end{eqnarray}
with the following background-dependent coefficients
\begin{eqnarray}
  C_{\varphi\varphi} &=& \frac{3{\dot{\varphi}}^2}{M_{\rm P}}-\frac{{\dot{\varphi}}^2 {\dot{\chi}}^2}{2{M_{\rm P}}^4 H^2}-
  \frac{{\dot{\varphi}}^4}{2{M_{\rm P}}^4 H^2}+\frac{2\dot{\varphi} V_{\varphi}}{{M_{\rm P}}^2 H} \\
   C_{\varphi\chi}&=&\frac{3\dot{\varphi}\dot{\chi}}{M_{\rm P}^2}-
   \frac{\dot{\varphi}\dot{\chi}^3}{2M_{\rm P}^4 H^2}-\frac{\dot{\varphi}^3\dot{\chi}}{2M_{\rm P}^4 H^2}+
   \frac{\dot{\varphi}V_{\chi}}{M_{\rm P}^2 H}+\frac{\dot{\chi}V_{\varphi}}{M_{\rm P}^2 H}+V_{\varphi\chi}  \\
   C_{\chi\chi}&=&\frac{3\dot{\chi}^2}{M_{\rm P}^2}-\frac{\dot{\chi}^4}{2M_{\rm P}^4H^2}-\frac{\dot{\varphi}^2\dot{\chi}^2}{2M_{\rm P}^4 H^2}+
   \frac{\dot{\chi}V_{\chi}}{M_{\rm P}^2 H}+V_{\chi\chi}  \\
   C_{\chi\varphi}&=&\frac{3\dot{\varphi}\dot{\chi}}{M_{\rm P}^2}-\frac{\dot{\varphi}\dot{\chi}^3}{2M_{\rm P}^4
   H^2}-\frac{\dot{\varphi}^3\dot{\chi}}{2M_{\rm P}^4 H^2}+\frac{\dot{\varphi}V_{\chi}}{M_{\rm P}^2
   H}+\frac{\dot{\chi}V_{\varphi}}{M_{\rm P}^2 H}+V_{\varphi\chi}
\; .
\end{eqnarray}
Following \cite{Gordon:2000hv}, we decompose the perturbations along
and perpendicular to the trajectory in the (homogeneous) field
space. The projection parallel to the trajectory is called the
instantaneous curvature, or adiabatic, perturbation whereas the one
orthogonal to the trajectory is termed the instantaneous
isocurvature, or entropy, perturbation. The velocity in the field
space is $\dot{\sigma} \equiv \sqrt{\dot{\varphi}^2+\dot{\chi}^2}$
and we can define the polar angle in the field space as
\begin{equation}\label{theta}
 \cos\theta\equiv\dot{\varphi}/\dot{\sigma}
\end{equation}
It is now useful to define the following Mukhanov-Sasaki variable:
\begin{equation}\label{Qsigma}
Q_{\sigma}=\cos\theta \, Q_{\varphi}+\sin \theta \, Q_{\chi} \; .
\end{equation}
In the flat gauge, $Q_{\sigma}$ represent the field perturbations
along the velocity in the field space. $Q_{\sigma}$ is also related
to the commonly used curvature perturbation, $\mathcal{R}$, of the
comoving hypersurface via
\begin{equation}\label{R}
{\mathcal R}=\frac{H}{\dot{\sigma}}Q_{\sigma} \; .
\end{equation}
Similarly the isocurvature perturbation is:
\begin{equation}\label{s-Q}
\delta s= -\sin \theta \, Q_{\varphi}+\cos\theta \, Q_{\chi} \; .
\end{equation}
It describes field perturbation perpendicular to the field velocity
in the field space and, by analogy with $\mathcal{R}$, we can define
a rescaled entropy perturbation, $\mathcal{S}$, through
\begin{equation}\label{S}
\mathcal{S} = \frac{H}{\dot{\sigma}}\delta s \; .
\end{equation}
The transformations described above basically amount to introducing
a new orthonormal basis in the field space, defined by vectors
\begin{eqnarray}\label{vectors}
E_{\sigma}&=& (E_{\sigma}^\varphi,E_{\sigma}^\chi) = (\cos\theta,\sin \theta) \; ,\\
E_s&=&  (E_s^\varphi,E_s^\chi) = (-\sin\theta, \cos\theta) \; ,
\end{eqnarray}
which turn out to be useful to express various derivatives of the
potential with respect to the curvature and isocurvature
perturbations. Employing an implicit summation over the indices $I,J
\in \{\varphi,\chi\}$, one thus finds
\begin{equation}\label{VI}
V_{\sigma}=E^I_\sigma V_{I} \; , \qquad V_{s}=E_s^I V_I \; ,
\end{equation}
and
\begin{eqnarray}\label{VII}
V_{\sigma\sigma}=E^I_{\sigma}E^J_{\sigma}V_{IJ}\; , \qquad V_{\sigma
s}=E^I_{\sigma}E^J_{s}V_{IJ}\; , \qquad V_{ss}=E^I_{s}E^J_{s}V_{IJ}
\; .
\end{eqnarray}
for the first and second derivatives.

By combining the Klein-Gordon equations for the background scalar
fields, eqs.~\eqref{eom-phi} and \eqref{eom-chi}, one obtains the
background EOMs along the curvature and isocurvature directions
\begin{eqnarray}\label{BG-curv-iso}
\frac{\mathrm{d}\dot{\sigma}}{\mathrm{d}t}+3H\dot{\sigma}+V_{\sigma}=0,\\
\dot{\theta}=-\frac{V_s}{\dot{\sigma}}.
\end{eqnarray}
With help of these one can show that the EOMs for curvature and
isocurvature perturbations become
\begin{equation}\label{perturbations}
\ddot{Q}_{\sigma}+3H\dot{Q}_{\sigma}+\left(\frac{k^2}{a^2}+C_{\sigma\sigma}\right)Q_{\sigma}+
\frac{2V_s}{\dot{\sigma}}\dot{\delta s}+ C_{\sigma s}\delta s =0,
\end{equation}
\begin{equation} \label{perturbationsDelSig}
\ddot{\delta s}+3H\dot{\delta
s}+\left(\frac{k^2}{a^2}+C_{ss}\right)\delta
s-\frac{2V_s}{\dot{\sigma}}\dot{Q}_{\sigma}+C_{s\sigma}Q_{\sigma} =
0,
\end{equation}
with coefficients given by
\begin{eqnarray}
  C_{\sigma\sigma} &=& V_{\sigma\sigma}-{\left(\frac{V_s}{\dot{\sigma}}\right)}^2
  +\frac{2\dot{\sigma}V_{\sigma}}{M_{\rm P}^2 H}
  +\frac{3{\dot{\sigma}}^2}{M_{\rm P}^2}-\frac{{\dot{\sigma}}^4}{M_{\rm P}^4 H^2}\\
  C_{\sigma s} &=& 6 H \frac{V_s}{\dot{\sigma}}+\frac{2V_{\sigma} V_s}{{\dot{\sigma}}^2}+2V_{\sigma s}+
  \frac{\dot{\sigma} V_s}{M_{\rm P}^2 H} \\
  C_{ss} &=& V_{ss}- {\left(\frac{V_s}{\dot{\sigma}}\right)}^2\\
  C_{s\sigma} &=&-6H\frac{V_s}{\dot{\sigma}}-\frac{2
  V_{\sigma}V_s}{{\dot{\sigma}}^2}+\frac{\dot{\sigma} V_s}{M_{\rm P}^2 H} \; .
\end{eqnarray}
A solution to these two coupled differential equations determines
the metric perturbation $\Phi$, which, in longitudinal gauge, is
related to the comoving energy density
\begin{equation}\label{epsilonm}
\epsilon_m=\dot{\sigma}\dot{Q}_{\sigma}+\left(3H+\frac{\dot{H}}{H}\right)\dot{\sigma}
Q_{\sigma}+V_{\sigma}Q_{\sigma}+2V_s \delta s
\end{equation}
via the Poisson-like relation
\begin{equation}\label{phi-epsilon}
\frac{k^2}{a^2}\Phi=-\frac{1}{2M_{\rm P}}\epsilon_m \; .
\end{equation}

The power spectra of curvature (adiabatic) and isocurvature
(entropy) perturbations are defined, respectively, as
\begin{equation}\label{power-spectra}
\mathcal{P}_{\sigma}(k)=\frac{k^3}{2\pi^2}\left\langle Q_{\sigma
\mathbf{k}}^{\star} Q_{\sigma \mathbf{k}'}\right\rangle
\delta^3(\mathbf{k}-\mathbf{k}') \; , \qquad \mathcal{P}_{\delta
s}(k)=\frac{k^3}{2\pi^2}\left\langle \delta s_{\mathbf{k}}^{\star}
\delta s_{\mathbf{k}'}\right\rangle \delta^3(\mathbf{k}-\mathbf{k}')
\; .
\end{equation}
The curvature and isocurvature perturbations are then evolved by
assuming initially, at conformal time $\tau_i$, a Bunch-Davies
vacuum. Therefore, when the wavelength of the two types of
perturbations is initially much smaller than the Hubble radius,
$k\gg aH$, we impose the initial conditions
\begin{equation}\label{initial conditions}
Q_{\sigma}(\tau_i)=\frac{e^{-ik\tau_i}}{a(\tau_i)\sqrt{2k}} \; ,
\qquad {\rm and} \qquad \delta
s(\tau_i)=\frac{e^{-ik\tau_i}}{a(\tau_i)\sqrt{2k}} \; .
\end{equation}
Inside the horizon these two modes are independent, because their
corresponding EOMs, eqs.~\eqref{perturbations} and
\eqref{perturbationsDelSig}, are independent in the limit $k \gg
aH$. However, as we will see in detail later, this does not hold
when the modes leave the horizon
\cite{Tsujikawa:2002qx,Byrnes:2006fr}.

Finally, let us introduce the two field slow-roll parameters
\begin{equation}\label{slow-roll-par}
\epsilon_{\varphi\varphi}=\frac{{\dot{\phi}}^2}{2M_{\rm P}H^2} \; ,
\qquad \epsilon_{\varphi\chi}=\frac{\dot{\phi}\dot{\chi}}{2M_{\rm
P}H^2} \; , \qquad \epsilon_{\chi\chi}=\frac{{\dot{\chi}}^2}{2M_{\rm
P}H^2}
\end{equation}
\begin{equation}\label{eta}
\eta_{IJ}=\frac{V_{IJ}}{3H^2}
\end{equation}
and
\begin{equation}\label{epsil}
\epsilon=\epsilon_{\varphi\varphi}+\epsilon_{\chi\chi}=-\frac{\dot{H}}{H^2}
\; .
\end{equation}

\end{document}